\documentclass[sigconf]{acmart}
\usepackage{float}
\usepackage{balance}

\AtBeginDocument{%
  \providecommand\BibTeX{{%
    \normalfont B\kern-0.5em{\scshape i\kern-0.25em b}\kern-0.8em\TeX}}}

\copyrightyear{2021} 
\acmYear{2021} 
\setcopyright{acmcopyright}\acmConference[CIKM '21]{Proceedings of the 30th ACM International Conference on Information and Knowledge Management}{November 1--5, 2021}{Virtual Event, QLD, Australia}
\acmBooktitle{Proceedings of the 30th ACM International Conference on Information and Knowledge Management (CIKM '21), November 1--5, 2021, Virtual Event, QLD, Australia}
\acmPrice{15.00}
\acmDOI{10.1145/3459637.3481979}
\acmISBN{978-1-4503-8446-9/21/11}

\begin{document}

\title{TagPick: A System for Bridging Micro-Video Hashtags and E-commerce Categories}

\author{Li He}
\affiliation{%
  \institution{University of Technology Sydney}
  \city{Sydney}
  \country{Australia}}
\email{li.he-1@student.uts.edu.au}

\author{Dingxian Wang}
\affiliation{%
  \institution{eBay Research America}
  \city{Seattle}
  \country{United States}}
\email{diwang@ebay.com}

\author{Hanzhang Wang}
\affiliation{%
  \institution{eBay Research America}
  \city{Seattle}
  \country{United States}}
\email{hanzwang@ebay.com}

\author{Hongxu Chen}
\affiliation{%
  \institution{University of Technology Sydney}
  \city{Sydney}
  \country{Australia}}
\email{hongxu.chen@uts.edu.au}

\author{Guandong Xu}
\affiliation{%
  \institution{University of Technology Sydney}
  \city{Sydney}
  \country{Australia}}
\email{guandong.xu@uts.edu.au}

\renewcommand{\shortauthors}{Li and Dingxian, et al.}

\begin{abstract}
Hashtag, a product of user tagging behavior, which can well describe the semantics of the user-generated content personally over social network applications, e.g., the recently popular micro-videos. Hashtags have been widely used to facilitate various micro-video retrieval scenarios, such as search engine and categorization. In order to leverage hashtags on micro-media platform for effective e-commerce marketing campaign, there is a demand from e-commerce industry to develop a mapping algorithm bridging its categories and micro-video hashtags. In this demo paper, we therefore proposed a novel solution called \textbf{TagPick} that incorporates clues from all user behavior metadata (hashtags, interactions, multimedia information) as well as relational data (graph-based network) into a unified system to reveal the correlation between e-commerce categories and hashtags in industrial scenarios. In particular, we provide a tag-level popularity strategy to recommend the relevant hashtags for e-Commerce platform (e.g., eBay).
\end{abstract}




\maketitle
\section{Introduction}
The increasing penetration and rapid development of social media have significantly shaped human's lifestyle nowadays. Various online multimedia contents occupy most of our spare time through the wide-spreading micro-video sharing platforms, e.g., Tiktok and Kuaishou~\footnote{https://en.wikipedia.org/wiki/Kuaishou}, for socialising, sharing and advertising as micro-video. These videos are compact and rich in multimedia content with multi-modalities, i.e., textual, visual, as well as acoustic information~\cite{wei2019mmgcn}.
Under such circumstances, the so-called social e-commerce, which allows us to make a purchase from a third-party merchant within the dedicated software, becomes one of the most popular means for online shopping. Many e-commerce providers (e.g., eBay and Taobao) start investing in such collaborated social platforms, and explore the drainage strategies on multimedia social platforms to increase the traffic for advertising efficiently~\cite{Gao2020}.

Micro-videos on these social platforms are usually associated with hashtags, which are commonly used to annotate the content aspect of the micro-videos and attract use attention ~\cite{cao2020hashtag}. Hashtag can be expressed by any arbitrary combination of characters led by a hash symbol ‘\#’ (e.g., \#beautifullife and \#heathystyle). Hashtags are created by users, they hence can be treated as the self-expression of users, conveying the users’ preferences on posts and their usage styles of hashtags. With these hashtags, users can easily search and manage their historical posts and track others’ posts. The wide and fast dissemination of above tagging behavior has made social media an ideal source for reflecting individual preference and collective intelligence. Therefore, e-commerce companies (e.g., eBay) have realized the marketing values of hashtags for advertisements and utilized the search volumes for the drainage of social e-commerce.

Despite the recent advancement in hashtag generation for various multimedia content (e.g., rich textual, microblog, micro-video), existing methods mainly focus on personalized hashtag recommendation~\cite{Wei2019}. However, little of them models the characteristics of user behavior in social media platforms. Moreover, many e-commerce companies (e.g.,eBay) urgently desire a hashtag selection module of leveraging the activity analysis of user hash-tagging on micro-video platforms (e.g., Tiktok), which would be incorporated into their current advertising system when they intend to post the video advertisements on those platforms, and provide highly relevant hashtags effectively so that the advertisements placed on the micro-video platform can attract more attention of users. 

The majority of existing studies on generating hashtags are mainly based on representation learning for the tagged objects on semantic level, such as collaborative filtering, generative models and deep neural networks~\cite{2012On, 2015hashtag}. However, picking hashtags for e-commerce advertisements (e.g., micro-video ads), diverting to bit different purpose is non-trivial due to the following challenges: First, the hashtag used by individual often contains ambiguity. For example, the hashtag ``\#fitness'' may be annotated to totally different categories of ``dietary'' or ``sports'' (weight loss). Users indeed always have their own preferences on hashtag usages. Second, in the context of e-commerce, the hashtag to be selected needs to reflect the marketing trend in social networks, serving for the drainage purpose. Therefore, how to find accurate and trendy hashtags for e-commerce advertising is the key problem. Inspired by the rich information contained in user post and tagging behaviors on social networks, we propose to derive the correlation between hashtags and e-commerce categories by graph-based learning. The benefits of this are three-fold: 1) User-annotated hashtags may contain lexical clues linking to categorical information of the products. As hashtags are generated to annotate, categorize, and describe a post (e.g., micro-video), they well describe the textual characteristics of targets. 2) The scenarios of user tagging have rich contextual information, e.g., textual comments, multimedia contents and different interactions (e.g., ``click'', ``follow'' or ``like''), which can provide complementary information. 3) The post hashtags are inherently related to the individual preferences, and such collective folksonomy of user preference does reveal the popularity and trend of user interest. 

To address the aforementioned problems, our system is built upon a graph-based learning model for bridging micro-video hashtags and e-commerce categories, which can jointly learn hashtag semantics and user behavior information from social networks. Then, the learned representations are used to measure the similarity between hashtags and e-commerce categories. In order to well illustrate the mapping results, we provide a web-based interface that enables users to input keywords of interested e-commerce categories and the corresponding output hashtags will be visualized in dashboard interactively for business use. The backend mapping algorithm not only works out the mapping results but also serves as an efficient search engine with interactive database system that supports a mass graph data processing from social media platforms and e-commerce platforms. 

The main contributions of the demonstration are summarised as follows: 1) We develop a novel hashtag mapping system, namely \textbf{TagPick}, which considers the implicitly linked data between social networks and e-commerce platform, and pick out the appropriate hashtags that match the category. 2) The backend of \textbf{TagPick} is built upon on a graph-based learning framework that exploits both user behavior and hashtag semantics for modeling the correlation between micro-video hashtags and e-commerce categories. We conduct extensive model training experiments on eBay and Tiktok datasets and the proposed framework is successfully embedded in eBay's cross-platform advertising system. 3) To make it easy to use and showcase how the hashtags are mapped to categories, we design an interactive web-based dashboard to visualize the results based on the open source Elastic Search (ES)~\footnote{https://www.elastic.co/}.

\section{System Overview}
\begin{figure}
    \centering
    \captionsetup{font={small}}
    \includegraphics[scale=0.35]{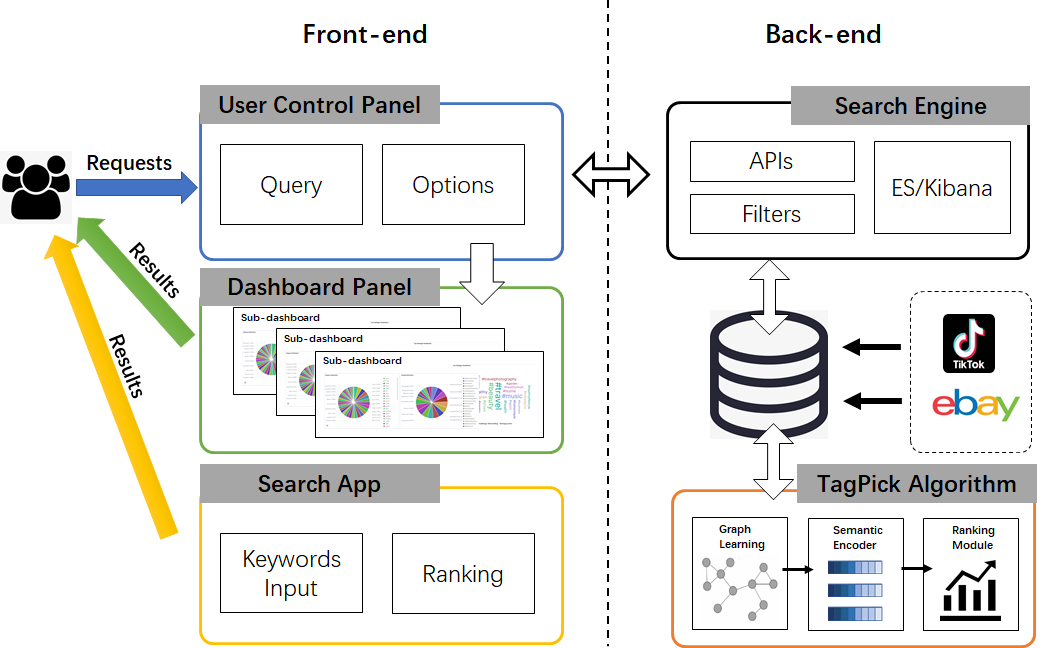}
    \caption{TagPick System Overview}
    \label{fig:overview}
    \vspace{-1em}
\end{figure}

The TagPick system consists of two major components as shown in Figure~\ref{fig:overview}: a web-based user interface and a backend which integrates a search engine and our hashtag bridging model. Specially, the web frontend consists of two components: 1) A user input interface to capture the entered keywords or trigger conditions. 2) A multi-modal dashboard to illustrate basic data statistics and mapping results. The backend mainly includes multiple modules: 1) A database to store the e-Commerce categories information and the pre-trained hashtags and its metadata. 2) The TagPick algorithm module based on graph-based deep learning. 3) An exploratory search engine that manages hashtag corpus, trending and related data distribution. The platform frontend, is comprised of a dashboards and a search interface, which are implemented using HTML, Javascript, and React, whereas the platform backend, including the algorithms, are written in Tensorflow with Python.

\paragraph{Frontend}
The implementation and development of the TagPick web platform (corresponding to the left part of Figure~\ref{fig:overview}). In Figure~\ref{fig:overview}, we can see a sample query and data visualization module being conducted on the TagPick platform. The three primary components in the figure are: 1) user controls modules, 2) dashboards, and 3) the search application. \textbf{User Control Modules}: with respect to the control module, the top part represents a user query (the drop-down menu provides several reference options); and the request from control panel applies only to the current dashboard. \textbf{Dashboards}: as mentioned above, the dashboard takes the input from user, and it can overlay and clear previous requests. In our system, we set up three sub-dashboards to illustrate top \(N\) hashtags, query results and time-aware trending. \textbf{Search Application}: to facilitate global search, this application provides search engine an interface and several functionalities, and will sort the results by default based on the ranking algorithm. In addition, the ranking tables contain additional information compared to the dashboard visualization.

\paragraph{Backend}
In this section, we describe the database, search engine and algorithmic design behind the TagPick algorithm (corresponding to the right part of Figure~\ref{fig:overview}). After getting the eBay categories and Tiktok hashtags datasets as input, the backend database stores these two different data corpuses separately and feeds the data to the TagPick algorithm. The algorithm module is the core of our platform, consisting of the following stages: 1) local graph structure around three types of entities - <hashtags, users, contents>, using a variation of deep walk with convolutional operations~\cite{deepwalk2014}; 2) the Tag2Vec model, which exploits several hierarchical relations such as hashtag-user, hashtag-content, hashtag-word, hashtag-category, and word-word to semantically understand the posted hashtags; 3) calculating the similarity score between hashtags and keywords (e.g., category name or product name), and ranking the results according to several statistical metrics (e.g., search volumes, post counts and trending).

\section{Technical Details}
The proposed TagPick algorithm encompasses three key algorithmic components: 1) The construction of user behavior analysis network based on graph model. 2) The semantic analysis module. 3) The similarity score calculation over multi-layered network.

\subsection{Graph Representation Learning}
Given multiple relations and entities in social networks, the goal of this module is to learn the user behavior containing the individual preference and the relevant category. Our proposed graph-based model mainly focuses on Graph Convolutional Networks (GCNs), which is a widely-used kind of graph models. Each layer of CGNs generates the intermediate embeddings by aggregating the information of items' neighbors. After stacking several GCN layers, we obtain the final embeddings, which integrate the entire receptive field of the targeted node~\cite{Kipf2017}. Specifically, on the constructed graph, we adopt the message-passing schema~\cite{2021autocite} to learn the users' preference on hashtags \(\mathbf{u}_i^h\) and contents \(\mathbf{u}_i^c\) based on a user \(u_i\)'s hashtag neighbors set \(\mathcal{H}_{i}\) and contents neighbors set \(\mathcal{V}_{i}\), respectively. After that, \(\mathbf{u}_i^h\) and \(\mathbf{u}_i^c\) are aggregated to represent the user preference \(u_i\). 

\textbf{User preference on hashtags} In our model, a user \(u_i\)'s preference on hashtags \(\mathbf{u}_i^h\) is modeled by aggregating the incoming messages from all the neighbor hashtags
\(\mathcal{H}_i\). According to the idea of message-passing, we transfer the message from a hashtag \(h_j \in \mathcal{H}_i\) to the user \(u_i\). As above, \(\mathbf{u}_{i}^{h}\) can be defined as:

\begin{equation}
\label{eq:1}
\mathbf{u}_{i}^{h}=\phi\left(\frac{1}{\left|\mathcal{H}_{i}\right|} \sum_{h_{j} \in \mathcal{H}_{i}} \mathbf{W}_{h}^{u} \mathbf{h}_{j}\right)
\end{equation}

where \(\mathbf{W}_{h}^{u}\) is the weight matrix which maps the hashtag vector into user embedding space. \(\phi(.)\) is the activation function and \(\left|\mathcal{H}_{i}\right|\) represents the number of neighbor hashtags.

\textbf{User preference on Content}
The user preference on the post content \(\mathbf{u}_i^c\) can be learned in the same way by aggregating the messages from all the neighbors' contents \(\mathcal{C}_i\). The passed message \(\mathbf{c}_k\) from one user's post represents all the contents in each post \(c_k\). Notice that a user's post content contains rich information, including text, a sequence of video clips and images. For example, the hashtags of a micro-video are usually provided based on the content. Therefore, the hashtags can better characterize the user preference on the post content. Similar to Eq.~\ref{eq:1}, the user preference on contents is the aggregation of the messages from all the neighbor contents.


\subsection{Semantic Encoder}
The e-commerce category and social hashtag contain lexical clues at different levels such as word-level and sentence-level, which provide different degrees of explainability of why these documents are inherently related.

\textbf{Word Encoder}
We learn the sentence representation via a bidirectional Recurrent Neural Network (RNN) with Gated Recurrent Units (GRU). The bidirectional GRU contains the forward GRU \(\overrightarrow{f}\) which reads sentence \(s_i\) from word \(w_1^i\) to \(w_{M_i}^{i}\) and a backward GRU \(\overleftarrow{f}\) which reads sentence \(s_i\) from \(w_{M_i}^{i}\) to \(w_1^i\). The sentence vector \(d^i\) is computed as bi-GRU processing. 

\textbf{Sentence Encoder}
Similar to word encoder, we use RNN with GRU to encode each sentence in news articles. Through the sentence encoder, we can learn the sentence representations. The annotation of sentence \(s_i\) is obtained by concatenating the forward and backward operations, which capture the context from neighbor sentences.

\subsection{Ranking on Multi-layered Network}
As discussed above, the user preference is obtained by combining the user preference on hashtags and post contents. Many combination methods can be applied here, such as concatenation and deep fusion models~\cite{Wang2020}. In this work, we tried the neural network-based fusion.

\textbf{Fusion Module} In this module, \(\mathbf{u}_{i}^{h}\) and \(\mathbf{u}_{i}^{c}\) are first concatenated and then fed into a fully connected layer to obtain the final representation of the user preference. Formally, the user preference is obtained by: 

\begin{equation}
\mathbf{u}_{i}=\phi\left(\mathbf{W}_{n n} Concat\left(\mathbf{u}_{i}^{v}, \mathbf{u}_{i}^{h}\right)+\mathbf{b}_{n n}\right)
\end{equation}

where Concat(.,.) is the concatenation operator; \(\mathbf{W}_{n n}\) and \(\mathbf{b}_{n n}\) indicate the learnable weight matrix and bias vector in the fully connected layer, respectively.

\textbf{Ranking Module}
Given a new e-commerce category keyword \(k_j\), the hashtags in \(\mathcal{H}\) could be arranged in the descending order of their similarity scores with respect to the category based on \(u_i\)'s preference. Specifically, the similarity score is computed by the dot product of \(\mathbf{u}_i^h\), \(\mathbf{u}_i^c\) and category keywords \(k_j\).

\section{Demonstration}
In this section, we show two scenarios of how TagPick can be used for mapping eBay category (i.e., which category to be advertised) and Tiktok micro-video hashtags.

\begin{figure*}
    \centering
    \includegraphics[scale=0.52]{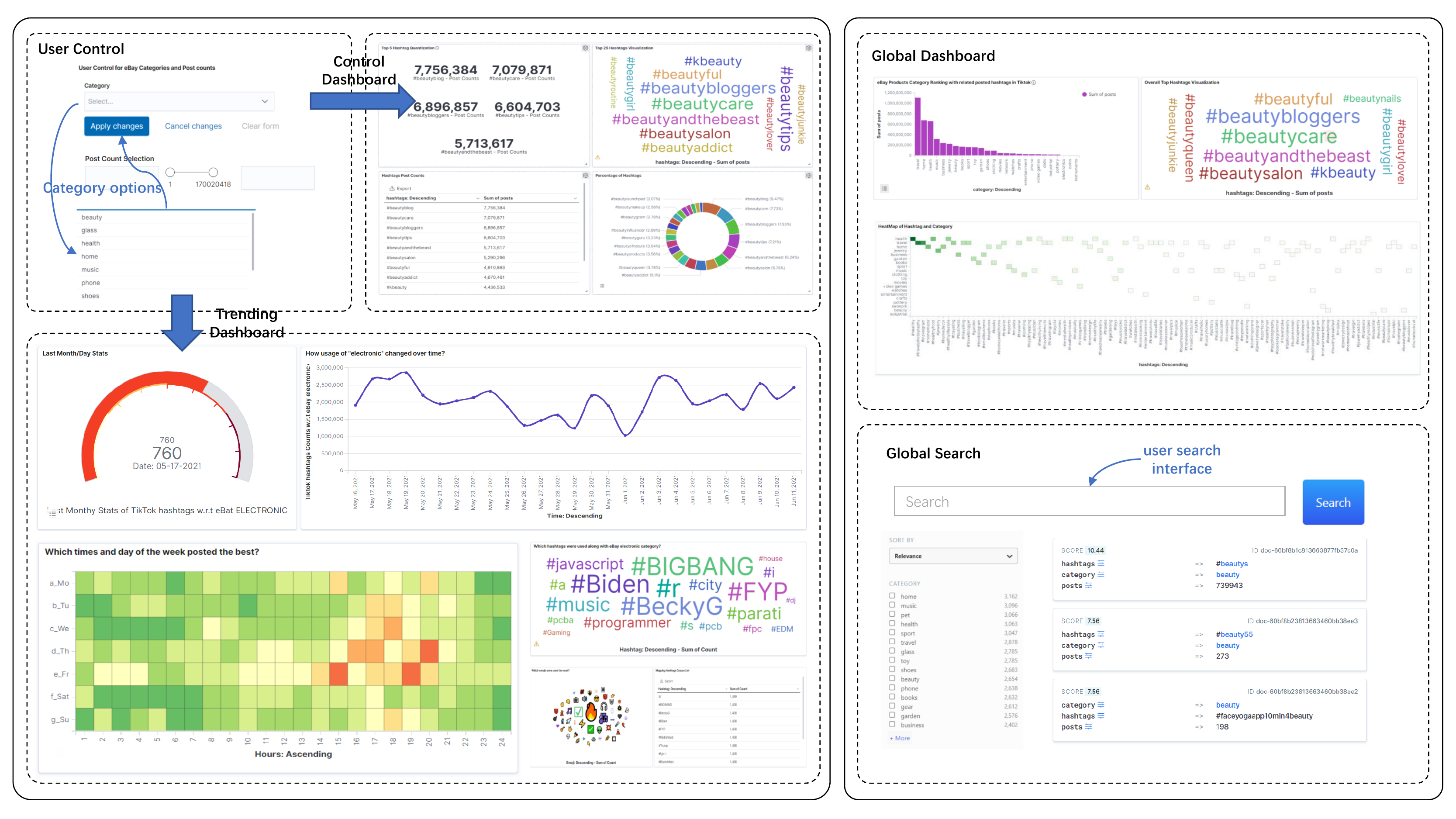}
    \captionsetup{font={small}}
    \caption{A Demonstration of TagPick: 1) User control panel and Dynamic dashboards (left); 2) Global static dashboard and global search user interface (right).}
    \vspace{-1em}
    \label{fig:demo}
\end{figure*}

\textbf{Local Control}
The frontend system consists control panels and dynamic dashboards. The control panel provides users with a multilevel relational menu to select eBay Categories from the database, as shown in Figure~\ref{fig:demo}. For example, if a user selects the eBay category keyword - "beauty", the range slider menu will provides filtering for different post counts of hashtags. We also provide a multi-level category retrieval interface, which can be dynamically added according to the contents of the database. Consider our platform is to be linked  to eBay advertising system, we provide a plug-in dashboard panel and support the export of retrieved result data as different file format (e.g., CSV~\footnote{https://en.wikipedia.org/wiki/Comma-separated\_values}). Users can select different category keywords and related hashtags by clicking the menu, the visual dashboard in the panel switches accordingly.

\textbf{Global Search}
Consider a user who wants to query the ranking list of hashtags associated with eBay categories, a user search interface (USI) is developed. The user first enters the keywords in the input box on the top of USI home page in bottom right of Figure~\ref{fig:demo}. The system would return several result panels which contain hashtag contents, similarity scores and indexes in database. These result panels are sorted by the score, illustrating results for category-level through a filter option. The higher the score, the more likely the hashtag matches the keyword in that category. The user can click each query result in the panel to check the details of the metadata, such as post counts, timestamp and other multimedia social content.

\section{CONCLUSION}
In this demo paper, we present TagPick, a fully-functional and easy-to-use platform for bridging e-commerce advertising and social network in industrial scenarios. TagPick leverages the distributed search engine as its data infrastructure, and adopts GCN-based methods as its core algorithm to perform our user behavior model training and calculate the similarity score over multi-layer encoders. TagPick provides a user-friendly Dashboard panel and USI for querying data and deploying it into their advertising system.

\section{Acknowledgments}
This work is supported in part by ARC under grants DP200101374 and LP170100891.
\bibliographystyle{ACM-Reference-Format}
\bibliography{demo_paper}
\balance
\end{document}